\documentclass[aps,pra,twocolumn,showpacs]{revtex4}
\usepackage{amsmath}
\usepackage{epsfig}
\usepackage{amssymb}
\usepackage{dcolumn}
\usepackage{graphicx}
\usepackage{dcolumn}

\usepackage{color}

\newcommand{\beq}{\begin{equation}}
\newcommand{\eeq}{\end{equation}}

\begin{document}
\date{\today}

\title{Nonlinear quantum piston for the controlled generation of vortex rings and soliton trains}

\author{Florian Pinsker}
\email{florian.pinsker@gmail.com}

\affiliation{Department of Applied Mathematics and 
Theoretical Physics, University of Cambridge, United Kingdom.} 

\author{Natalia G. Berloff}
\email{N.G.Berloff@damtp.cam.ac.uk}

\affiliation{Department of Applied Mathematics and 
Theoretical Physics, University of Cambridge, United Kingdom.}

\author{V\'{\i}ctor  M. P\'erez-Garc\'{\i}a}
\email{victor.perezgarcia@uclm.es}

\affiliation{Departamento de Matem\'aticas, Escuela T\'ecnica
Superior de Ingenieros Industriales, and Instituto de Matem\'atica Aplicada a la Ciencia y la Ingenier\'{\i}a (IMACI),
Universidad de Castilla-La Mancha, 13071 Ciudad Real, Spain.}

\begin{abstract}
We propose a simple way to generate nonlinear excitations in a controllable way by managing interactions in Bose-Einstein condensates.  Under the action of a quantum analogue of a classical piston the condensed atoms are pushed through the trap generating vortex rings in a fully three-dimensional condensates or soliton trains in quasi-one dimensional scenarios. The vortex rings form due to  transverse instability of the shock wave train enhanced and supported by the energy  transfer between waves. We elucidate in which sense the self-interactions within the atom cloud define the properties of generated vortex rings and soliton trains.  Based on the quantum piston scheme we study the behavior of two component  Bose-Einstein condensates and analyze how the presence of an additional superfluid influences the generation of vortex rings or solitons in the other component and vice versa. Finally, we show the dynamical  emergence of skyrmions within two component systems in the immiscible regime.
\end{abstract}


\maketitle

\section{Introduction} One of the most remarkable achievements in quantum physics in the last decade was that of Bose-Einstein condensation (BEC) in ultra-cold alkaline atomic gases. These physical systems have a high potential for supporting quantum nonlinear coherent excitations and many types of nonlinear waves have been experimentally observed \cite{dark,bright,gap,vector,v1a,v1b,v1c,vrings,vrings2,vrings3,vrings5,sh1,blowup} or theoretically proposed to exist (see e.g. the reviews \cite{Panos,Panos2}) in ultra-cold quantum degenerate gases.
The list includes dark solitons \cite{dark}, bright solitons \cite{bright}, bubbles \cite{yin} and gap solitons \cite{gap}, vector solitons \cite{vector}, vortices, vortex lattices and giant vortices \cite{pins}, vortex rings \cite{vrings,vrings2,vrings3}, dark ring-shaped waves \cite{vrings5}, shock waves  \cite{vrings2,sh1}, collapsing waves \cite{blowup} and many others.
In this way Bose-Einstein condensates (BECs)  are, apart from their fundamental role in quantum physics, exceptional physical systems for the manifestation and study of nonlinear phenomena and in particular due to their rather simple theoretical description.

Among the various nonlinear excitations the vortex ring occupies a special place. Vortex rings are essentially three-dimensional topological nontrivial structures appearing in either classical \cite{Batchelor} or quantum \cite{Donnelly} fluids. They are able to propagate in cylindrically trapped BECs as stable objects \cite{Komineas}, similarly to classical fluids \cite{JR}. This is an essential difference to most other solitonic structures that become unstable when passing to a fully three-dimensional setting, e.g. one dimensional bright solitons that are unstable to blow-up \cite{blowup} or dark solitons, that are unstable to the snake instability \cite{snakeInstability,berloffBarenghi,vrings}.  This leaves the vortex ring as the only dynamically nontrivial nonlinear excitation observed in fully three-dimensional BECs.

Vortex rings were first observed in BECs as the outcome of the decay of dark solitons \cite{vrings} and as a result of the decay of quantum shock waves \cite{vrings2}. More recently they have been  observed to appear  during complex oscillations in  soliton-vortex ring structures \cite{vrings3} and during the merging of BEC condensate fragments \cite{Scherer}. However,  generating vortex rings involved complicated nonlinear phenomena and in general
 a simple mechanism allowing the controlled generation of a prescribed finite number of vortex rings is still missing. The main purpose of this paper is to propose such a mechanism allowing the generation of a few vortex rings in a highly controllable way.   The same method can be used to create soliton trains of  certain frequencies within a one-dimensional model. We will also extend the concepts to coupled BECs showing how skyrmions can be generated using similar techniques.
 
 The plan of this paper is as follows. First in Section \ref{idea} we introduce the main physical idea of a nonlinear quantum piston. In
  Section \ref{equations} we introduce the mathematical equations and nondimensionalisations used throughout the paper.  Next we discuss the nonlinear excitations in the form of dark and bright soliton trains for one dimensional problems  for single (Sec. \ref{sec-dark}) and two-component (Sec. \ref{sec-dark2})
condensates. The controlled generation of vortex rings and skyrmions is discussed in Section \ref{sec-vortex}. Finally we summarize our conclusions in Section \ref{sec-conclusions}.

\section{Physical idea}
\label{idea}

 The process of vortex ring generation in classical fluids has received a substantial treatment in the literature. 
One of the most standard ways to obtain vortex rings in classical fluids involves moving a piston through a tube, resulting in a vortex ring being generated at the tube exit. A standard generation geometry consists of the tube exit mounted flush with a wall with the piston stroke ending at the tube exit \cite{Glezer}.

 In this paper we will use something conceptually much simpler utilizing the possibilities opened by space-dependent Feschbach resonance management in a BEC. Since the first achievements in scattering length control in BECs \cite{FB1}, the technique of Feschbach resonance management has been improved and used in many different applications. Presently, the level of control of the scattering length allows for its very precise tuning \cite{Hulet} and nothing prevents an extended control of the interactions leading to a space dependent scattering length. A large number of theoretical papers have  studied nonlinear phenomena in systems with managed interactions  (see e.g. Refs. \cite{Panos,V1,V2,V3,BadPan,SE1,SE2} and references therein).

 \begin{figure}
\epsfig{file=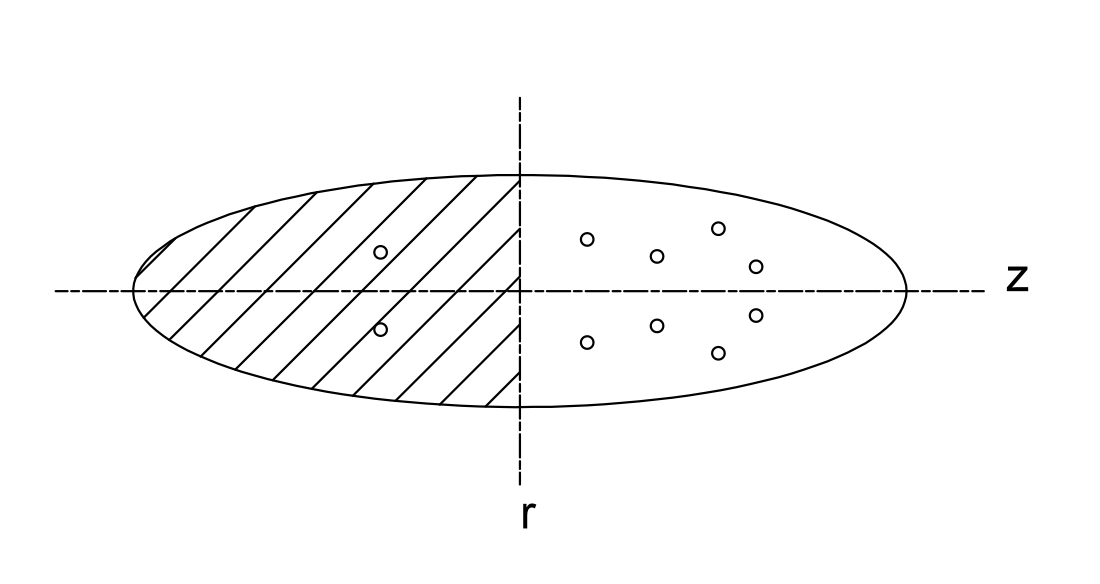,width=\columnwidth}
\caption{Schematic diagram of the quantum piston idea, where the ellipse symbolizes the trapped BEC. The trap is supposed to be radially symmetric and elongated along the $z$-axis. The area of change in self-interactions is illustrated by the hatched lines. The small circles represent the cross sections of possible vortex rings obtained as a result of the flow induced by the asymmetric interactions playing the role of a quantum piston. \label{post}}
\end{figure}
The physical idea is illustrated in Fig. $\ref{post}$. Starting from a single equilibrium BEC in a trap with a given value of the scattering length $a>0$, we propose modifying interactions in half of the space (say the left hand side $z<0$) to a larger value $a_L>a=a_R$  instantaneously. This change would affect the initial configuration by inducing the transverse expansion of the atomic cloud for $z<0$ generating a flow towards the region with a smaller interaction value
at $z>0$. This process is analogous to the piston-driven flow through an aperture used to generate vortex rings in classical fluids. These effects are achieved simultaneously with a single action on the interactions without restoring to complicated external potentials. We also consider a smooth change of scattering length that would be more realistic in experiments. Modifying interactions to be attractive $a<0$ on one side  would allow the atom cloud with repulsive self-interaction at $z>0$ to expand towards this region. The nature of nonlinear excitations generated in this way would be different to the repulsive case due to the difference in interatomic relations \cite{Pita2,nat}. Generating a controlled flow within a Bose gas with entirely attractive interactions by imposing a change in interactions would cause similar nonlinear excitations.

 Applying the concept of a quantum piston to a two component BEC enables new physics to come into play. One scenario we will consider in this paper is the case in which the  interactions are changed only in one of the components that  would generate a counterflow between two components thus creating
 excitations  involving the other component, for example generating skyrmions  that are stable vortices with the  second component filling in the core.

\section{Mathematical models}  
\label{equations}

We consider a single component BEC modeled in the mean field limit by the nonlinear Gross-Pitaevskii equation \cite{Pita,Gross} 
\begin{equation}\label{modeln}
 i \hbar \frac{\partial \psi}{\partial t} = - \frac{\hbar^2}{2 m}\Delta \psi + V^{\rm nat}_{\text{ext}}(r,z)\psi + g^{\rm nat}(z) |\psi|^2 \psi,
\end{equation} 
for particles of mass $m$ with self-interactions defined by $g^{\rm nat}(z) = 4 \pi \hbar^2 a_s(z)/m$, where $a_s(z)$ is the s-wave scattering length. The condensate wave function $\psi$ describing a condensate of $N$ bosons located in $\Omega \subset \mathbb R^3$ satisfies the mass constraint $\|\psi \|^2_{L_2(\Omega)}= N$. Confinement is due to an external elongated axisymmetric trap $V^{\rm nat}_{\text{ext}} = (m\omega^2 r^2 +m\omega^2_z z^2)/2$, where the frequencies  satisfy $\omega \gg\omega_z$. We use the rescaled GPE given by
\begin{equation}\label{model}
  i \frac{\partial \psi}{\partial t} = - \Delta \psi + V_{\text{ext}}(r,z)\psi + g (z) |\psi|^2 \psi.
\end{equation} 
Here the spatial coordinates are measured in the healing length of the transverse ground state $a_0=(\hbar/m \omega \sqrt{2})^{1/2}$ and time $t$ in $\sqrt{2}/\omega$, respectively, while the energies and
frequencies are measured in units of $\hbar \omega/\sqrt{2}$ and $\omega/\sqrt{2} $ respectively,   and $V_{\text{ext}} = (\lambda^2 r^2+\lambda_z^2z^2)/2$.  We rescale the wave function such that it is normalized to 1, i.e., $\|\psi \|^2_{L_2(\Omega)}= 1$. Then $g (z) =4\pi a_s(z) N/a_0$, which is
proportional to the local value of the s-wave scattering length $a_s(z)$ and will be taken, starting from $t=0^+$,  to be  the step-like function defined by
\begin{equation}\label{step}
g (z) = g^L +  \frac{g^R-g^L}{1 + e^{-2kz}},
\end{equation}
where the constituents of the local self-interaction are $g^R$ and $ g^L$ and the `smoothness' parameter is taken to be $k>0 $.  Finite $k$ accounts for a gradual change in scattering length across $z=0$. In the limiting case $k \to \infty$ the coupling parameter $g (z)$ becomes a step function
\begin{equation}
g(z) \stackrel{k \to \infty}{\longrightarrow} \begin{cases} g^R, & z>0 \\ (g^R + g^L)/2, &  z= 0 \\ g^L, & z<0. \end{cases}\qquad 
\label{g}
\end{equation}

In addition we will  study two component Bose-Einstein condensates within a similar quantum piston setting. The wave functions of component A and B will be assumed to be governed by a system of coupled nonlinear Schr\"odinger-type equations \cite{Pita2,Smith}. 
\begin{subequations}
\begin{multline}\label{modela}
 i \hbar  \frac{\partial \psi_A}{\partial t}  =  \left(- \frac{\hbar^2}{2 m_A}\Delta + V^{\rm nat}_{\text{ext},A}(r,z) + \right. \\  \left. + g^{\rm nat}_{A}|\psi_A|^2 + g^{\rm nat}_{AB} |\psi_B|^2 \right) \psi_A,
\end{multline}
\begin{multline}\label{modelb}
i \hbar  \frac{\partial \psi_B }{\partial t}  = \left(- \frac{\hbar^2}{2 m_B}\Delta + V^{\rm nat}_{\text{ext},B} (r,z)+ \right. \\ \left. g^{\rm nat}_{B} |\psi_B|^2 + g^{\rm nat}_{BA}|\psi_A|^2\right) \psi_B.
\end{multline} \label{sysDim}
\end{subequations}

Here the mass of a boson from component $i \in \{A,B\}$ is denoted $m_i$, the symbol $g^{\rm nat}_i$ refers to the corresponding self-interactions and we consider a scaling for which the normalization is $\|\psi_{i} \|^2_{L_2(\Omega)} = N_i$, where $N_i$ denotes the number of atoms of species $i$. Depending on the context self-interactions $g^{\rm nat}_i$ can be thought of either as being constant or step functions. Those step-functions are formally defined as in the single component case. The cross-interaction strengths are given by $g^{\rm nat}_{ij} = 2 \pi \hbar^2 a_{i j}/ m_{ij}$ with $i \neq j$ and the reduced mass given by $m_{ij} = m_i m_j/(m_i + m_j)$. In this paper we will take $a_{ij}=a_{ji}$ and  $m = m_A=m_B $.  %All interactions taking place between the particles are assumed to be repulsive, i.e., $g^{\rm nat}_{i} > 0$ and $g^{\rm nat}_{ij} > 0$ initially. }
%We note that a necessary condition for having phase separation is ${g^{\rm nat}_{AB}}^2 > g^{\rm nat}_A g^{\rm nat}_B$ (immiscible regime). 
The harmonic potentials are given by
$V_{\text{ext},i} = m_i (\omega r^2+\omega_z^2z^2)/2$, where the $\omega$'s denote the corresponding trapping frequencies. 

The non-dimensionless form  is obtained via the transformation $\boldsymbol{x} \to a_0 \boldsymbol{x}$, $t \to t \sqrt{2} /
 \omega$ and 
 $\psi_i \to \psi_i \left( \frac{1}{L_i} \right)^{1/2}/a_0$ with $\frac{1}{L_i} = g_i \hbar^2 /(2m g^{\rm nat}_{i})$ where 
 the nondimensionalized self-interaction strength $g_i$ has been introduced. In those terms our coupled system is given by
 \begin{subequations}\label{sys}
\begin{eqnarray}\label{A}
i   \frac{\partial \psi_A}{\partial t}  = \left( - \Delta + V_{\text{ext},A} + g_A |\psi_A|^2 + g_{AB} |\psi_B|^2 \right) \psi_A \\ \label{B}
  i   \frac{\partial \psi_B}{\partial t}   =  \left(- \Delta + V_{\text{ext},B} + g_{AB} |\psi_A|^2  + g_B |\psi_B|^2 \right) \psi_B
\end{eqnarray}
\end{subequations}
with  $ V_{\text{ext},i} =  \lambda_{i}^2 r^2 + \lambda_{z,i}^2 z^2$ and $g_{ij} =  g_j g^{\rm nat}_{ij}/g^{\rm nat}_j$.  We choose parameters such that the mass constraints $\|\psi_{i} \|^2_{L_2(\Omega)} = 1$, $\Omega \subset \mathbb R^3$.

To follow the time evolution of the fields $\psi_i$ numerically we have used a fourth order finite difference scheme in space together with a fourth order Runge-Kutta discretization in time.
 
In the next section we consider the generation of nonlinear excitations by changing the interaction strength on one half of the domain for quasi-one dimensional BECs.

 \section{Emergence of soliton trains in quasi-one dimensional single component BECs}  
\label{sec-dark}

 The starting point of our investigation of the controlled generation of nonlinear excitations is a strongly cigar shaped Bose-Einstein condensate, i.e., $\omega \gg \omega_z$. Neglecting trapping in $z$ by setting $\omega_z = 0$, the evolution equation \eqref{model} for the wave function becomes \cite{marko}
\begin{equation}\label{model1d}
 i  \frac{\partial}{\partial t}  \psi_{\text 1D}= - \frac{\partial^2}{\partial z^2} \psi_{\text 1D}  + g (z) |\psi_{\text 1D}|^2 \psi_{\text 1D} - \mu\psi_{\text 1D},
\end{equation} 
where we have introduced a chemical potential via $\psi_{\text 1D}= \psi_{\text 1D} e^{-i \mu t}$, dropped the mass constraint as we consider an infinitely spread BEC here and rescaled time. Initially, i.e., for $t < 0$ the  single coherent condensate is uniformly distributed and lies at rest. By $n_0 = \mu /g$ we denote the associated constant equilibrium density distribution and the corresponding self-interaction strength by $g$. To see the effect  of changing self interactions  at $t=0$ instantaneously, i.e., $g \to g(z)$, on the initial state (such that we have step-like self interactions $g(z)$ for all times $t > 0$) we have simulated the dynamical behavior of $\psi$  governed by  \eqref{model1d}  for different parameter combinations of $g(z)$ \cite{mynote}.

 \begin{figure}
\epsfig{file=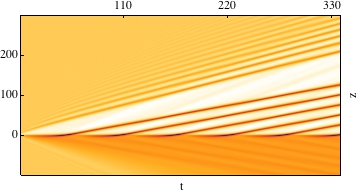,width=\columnwidth }
\caption{ Pseudo-color density plot $\rho(z)=\left|\psi_{\rm 1 D}  (z,t) \right|^2$ of a uniformly distributed Bose gas with constant interaction $g =1=g^R$ at $t=0$ evolving in time as a change of $g^L/g^R =3.4$ has been implemented for $t>0$. The change in interactions is sharp at $z=0$, i.e., $k \to \infty$ in $\eqref{step}$. Here luminosity is proportional to density. The dimensionless units are used as specified in the main text \label{rain}.}
\end{figure}

\subsection{Results}

Starting with a uniformly distributed Bose gas with repulsive self-interactions set to $g=1$ we observe that for a moderate increase in the interaction strength on the left-hand side the outflow does not produce any solitary trains.  When the spatial change in self-interactions is sufficiently large, specifically $g^L/g^R > 2.2$  the transport of atoms from the region of higher interactions on the left to the one of lower interactions on the right is accompanied by the emergence of  a dark soliton train. 

%Fig.~\ref{rain} shows the time snapshots of the density plots of a BEC for numerically computed states $\psi_{\text 1D}$ at time $t=0$ (dotted line) for constant $g$ and at $t=47.25$ (dashed line) and $t=182.25$ (solid line) due to the for $t > 0$ imposed step-like change in self-interactions $g(z)$ (in the limit $k \to \infty$). The density profiles at $t=47.25$ and $t=182.25$ represent the early stages of dark soliton train generation. {\blue The figure illustrates the formation of the leading solitary wave (dashed line) moving to the r.h.s. that initiates the solitary train (solid line) following in the same direction. As the perturbation extends further into the region a new wave density dip is formed and so on. 

 The wave generated when the interactions are increased on the left
half of the cloud  ($z<0$) via $g \to g(z)$  leads  to a formation of dispersive shock that
propagates on the background density that sets the reference sound
speed. As shown by many authors \cite{k1,k2,k3,k4,k5} the 1D shock profile can be  found
by matching the high- and low-intensity boundaries. In such a shock
the inner (slow) nonlinear part of the front is  a train of dark or
gray solitons, while the outer (fast) part is a low-intensity region
with oscillations that are effectively sound-like \cite{k1,k2,k3,k4,k5}. As the fast
outer part propagates further into  the less interactive region, the inner
part adopts more and more pulses in the solitary train, which is
clearly seen on Fig.~\ref{rain}. Subsequently the flow of the condensate manifests itself as a regular array of density depletions moving at a constant speed while keeping the shape over time.  Results in this regime also agree well with simulations of the nonpolinomial Schr\"odinger equation in trapped systems reported in Ref. \cite{BadPan}.  In addition we note that in a different context soliton patterns arise due to a mechanism where two spatially distinct condensates collide within a harmonic trap  \cite{scker} and thereby generate nonlinear excitations. The two condensates in this case have different global phases, so joining
the two condensate together is analogous to phase imprinting in a
single condensate \cite{mat}. In our case we have a condensate with the same global phase
where vortices and solitary trains are formed  {\it dynamically}.
 \begin{figure}
\epsfig{file=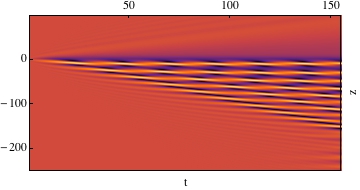,width=\columnwidth }
\caption{ Pseudo-color density plot $\rho(z)=\left|\psi_{\rm 1 D}  (z,t) \right|^2$ of a uniformly distributed Bose gas with constant interaction $g =1=g^R$ at $t=0$ evolving in time as a change of $g^L/g^R =-1$ has been implemented for $t>0$. The change in interactions is sharp at $z=0$, i.e., $k \to \infty$ in $\eqref{step}$. Here luminosity is proportional to density. The dimensionless units are used as specified in the main text. \label{grain}}
\end{figure}

 In Fig.~\ref{grain} we show the  evolution for a BEC between $t=0$ and $t=160$ in the case where we have  changed self-interactions to negative (attractive) values at the l.h.s. In this particular example parameters have been chosen to be $g=1=g^R$ and $g^L = - 1$. The emerging structure can be identified as  a bright soliton train appearing in a similar process of formation of individual solitons in a localized reservoir. \cite{SE1,SE2}.  Furthermore  it can be seen in Fig.~\ref{grain} that bright solitons  remain approximately at the same position. Similarly to the formation of a dark soliton train as an initial shock wave propagates a bright soliton train is generated.
 %Switching one half of an attractive BEC to repulsive interactions leads to a similar generation of bright soliton trains.
  \begin{figure}
 \epsfig{file=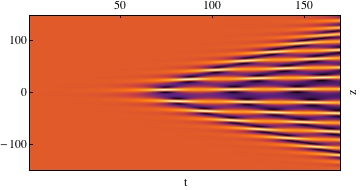,width=\columnwidth }
\caption{ Pseudo-color density plot $\rho(z)=\left|\psi_{\rm 1 D}  (z,t) \right|^2$ of a uniformly distributed Bose gas with constant attractive interaction $g =-1=g^R$ at $t=0$ evolving in time as a change of $g^L/g^R =0.99$ has been implemented for $t>0$. The change in interactions is sharp at $z=0$, i.e., $k \to \infty$ in $\eqref{step}$. Here luminosity is proportional to density. The dimensionless units are used as specified in the main text. \label{brain2}}
\end{figure}

 Starting  with a system of attractively interacting atoms  the introduction of even small change in interactions leads to the generation of a bright soliton train as Fig. \ref{brain2}  illustrates  for the case of  $g=-1=g^R$ and $g^L/g^R = 0.99$. Unlike the cases involving repulsive interactions this bright soliton train slowly expands in both directions. However, as $g^L/g^R \to 0$ bright solitons can only be observed for $z>0$.

\subsection{Smooth vs. abrupt change in self-interactions} 

In real experiments one would expect a more gradual change of the interaction strength across $z=0$. Thus to describe more realistic situations  a finite (though maybe small) $k$ has to be considered within Eq.  \eqref{step}. 

\begin{figure}
\epsfig{file=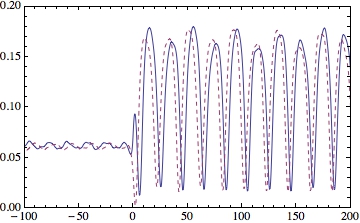,width=\columnwidth}\put(-252,130){ \textcolor{black}{$\rho(z)$}}\put(-108,2){ \textcolor{black}{$z$}}
\caption{Density plot $\rho(z)=\left|\psi_{\rm 1 D}  (z,t) \right|^2$ of numerically computed density profiles  at $t=510$ for  $g^L/g^R = 4$ and $g^R=1$ and  $k \rightarrow \infty$ (solid line) and  $k=1.5$ (dashed line). The dimensionless units are used as specified in the main text
 \label{lab}}
\end{figure}

In several series of simulations for attractive as well as repulsive condensates we have observed the same qualitative dynamics as for the step-function case, the structure of the solitary wave train and the threshold for its appearance has been very similar even for small $k$, i.e., a very smooth step. In Fig. \ref{lab} we present an example where $k=1.5$ showing the very small deviation in emerging soliton trains due to sharp and gradual changes in self-interactions. Hence, for simplicity we will turn our attention on the limit  $k \to \infty$ in what follows.

 \begin{figure}
\epsfig{file=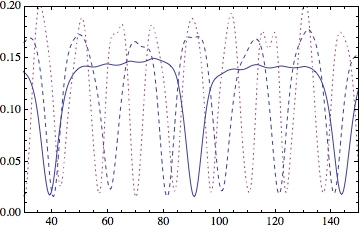,width=\columnwidth}\put(-252,135){ \textcolor{black}{$\rho(z)$}}\put(-120,3){ \textcolor{black}{$z$}}
\caption{ Density plots $\rho(z)=\left|\psi_{\rm 1 D}  (z,t) \right|^2$ of numerically computed density profiles  with step function interactions ($k \rightarrow \infty$) at time $t = 500$  of an initially uniformly distributed Bose gas with $g =1= g^R$ for different self interaction strenghts ratios $g^L/g^R = 2.5$ (solid line), $g^L/g^R = 4$ (dashed line) and $g^L/g^R = 6$ (dotted line). The dimensionless units are used as specified in the main text \label{freq}}
\end{figure}

 \begin{figure}
\epsfig{file=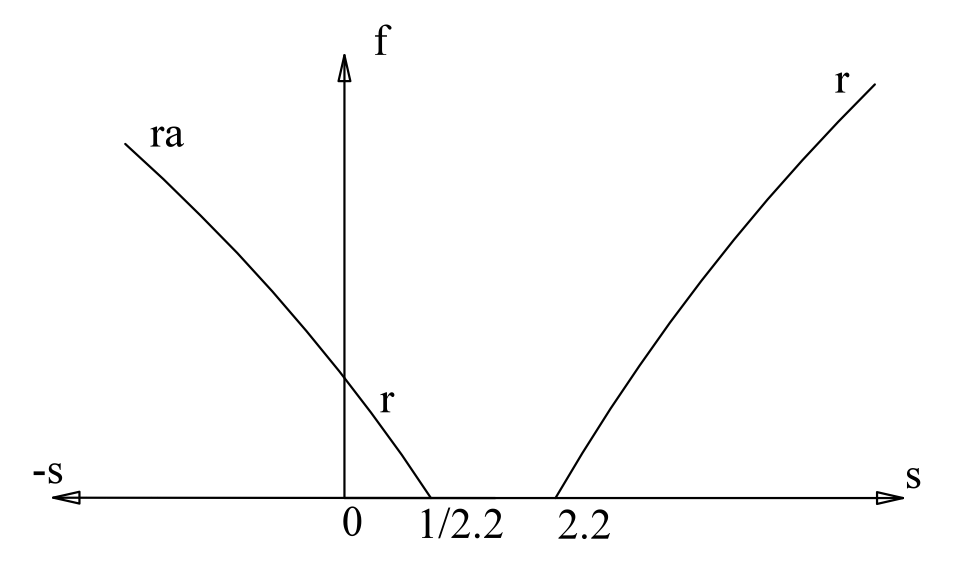,width=\columnwidth}
\caption{A schematic diagram of the relationship between the frequencies $f$ of the soliton trains and the change in interactions $s= g^L/g^R$.The dimensionless units are used as specified in the main text  \label{freq3}}
\end{figure}

\subsection{Properties of soliton trains}

For repulsive interactions one would expect that an increase in the change of interactions, $g^L/g^R$, produces larger flow, so leads to an increase in the spatial frequency, i.e.,  the number of solitons within some fixed space interval.
An example of this behavior is shown in Fig. $\ref{freq}$ where an initial distribution specified by $g=1=g^R$ changes its profile when the interactions are set to  $g^L/g^R = 2.5$, $g^L/g^R = 4$ and $g^L/g^R = 6$. The final profiles for $t=500$ are shown. Fig. $\ref{freq}$ shows that larger asymmetries in the interaction lead to higher spatial frequencies. In addition we observe that for large asymmetries in interactions  the wavefunction profiles resemble the square of a sinus function while for smaller asymmetries the profiles of the density depletions resemble an array of squares of hyperbolic tangents near their minima, i.e., there is a qualitative difference in the form of the density profile depending on the imposed change in interactions. We have also studied the velocities of generated dark soliton trains and found them to be almost independent on the change in interaction strength. 

 The numerically obtained relationship between the frequencies of soliton trains and the change in interactions $g^L/g^R\in [-2,8]$ is given in Fig. \ref{freq3} for  $g =1$. Here $r$ denotes entirely repulsive BEC and $ra$ a condensate where we switched to attractive values on one side. For entirely attractive interactions one finds that changing self-interactions to different values leads to bright soliton trains with in general different  frequencies on both sides. 

\subsection{Analytical approximations to the soliton train profiles}

Eq. (\ref{model1d}) with spatially dependent interactions does not admit in general exact analytical solutions. While we will be constructing 
 solutions of the full equation in Appendix \ref{solutions2} (for step-function like self-interactions in the static case), for now we will restrict our attention to the construction of phenomenological solutions fitting the dynamics for uniform $g$ that can be expected to approximate the solutions far from $z=0$ (the point where the nonlinearity has the transition from $g^L$ to $g^R$).
First we consider the repulsive and then the attractive case. The differential equation \eqref{model1d} for constant $g$ is  integrable \cite{Beth2} and the solution representing a single dark soliton \cite{Tsuzuki} for $g > 0$ can be written as
\begin{multline}\label{tanh}
\psi_{g>0}(z,t)  = \sqrt{n_0} \Bigg[i  \frac{v}{c } + \sqrt{ 1- \left(\frac{v}{c}\right)^2 }\cdot \\  \cdot \tanh \Bigg(   \sqrt{ 1- \left(\frac{v}{c}\right)^2 } \cdot \sqrt{\frac{n_0 g}{2}} \cdot \left(z -  v t \right) \Bigg)  \Bigg].
\end{multline}
Here $v$ is the velocity of the soliton, $c=\sqrt{2 n_0 g}$ is the Bogoliubov speed of sound, $n_0$ denotes the equilibrium one particle density distribution and one sets $\mu = n_0 g$. In order to get a periodic solution describing a dark soliton train we interchange the hyperbolic tangens in $\eqref{tanh}$ with a Jacobi elliptic function by letting $\tanh \to \operatorname{ sn } $,
\begin{multline}\label{appr}
\psi_{\rm dt}(z,t)  = \sqrt{n_0} \Bigg[i  \frac{v}{c} + \sqrt{1- \left(\frac{v}{c}\right)^2  }\cdot \\  \cdot \operatorname{ sn } \Bigg(   \sqrt{ 1- \left(\frac{v}{c}\right)^2 } \cdot \sqrt{\frac{n_0 g}{2 p^2 }} \cdot \left(z -    v t \right) \bigg | p^2 \Bigg)  \Bigg].
\end{multline}
A straightforward calculation shows that it is approximately solving $\eqref{model1d}$, if $\mu$ is chosen to be $\mu= g n_0 \left( \frac{(1+p^2)}{(2 p^2)}(1- \frac{v^2}{2 n_0 g}) +  \frac{v^2}{2n_0 g} \right)$. 
Indeed, the above approximate solution becomes exact either if  $p \rightarrow 1$, where  $\operatorname{ sn }(z|1) = \tanh(z)$,  or if $v \rightarrow 0$, so it generalizes the single soliton expression $\eqref{tanh}$. It is known that Jacobi elliptic functions are periodic solutions of the GP equation with repulsive and attractive interactions  \cite{kuz,hand,Carr,Carr2,Zhong}. The sinus amplitudinis  interpolates between a trigonometric and a hyperbolic function and its dependency on each is controlled by the real-valued elliptic modulus $p \in [0,1]$.

 One property of the analytic dark soliton train $\eqref{appr}$ is that its density profile fits with the computationally obtained profile, iff the self-interaction strength $g$ of the analytical solution equals the one used for generating the numerical solution.  This in turn enables us to deduce the effective interaction strength $g$ between the condensed atoms from the form of the numerically generated dark soliton train, i.e., from the density profile of the atom cloud by means of $\eqref{appr}$. For details we refer to the Appendix \ref{ApX}.

\begin{figure}
\epsfig{file=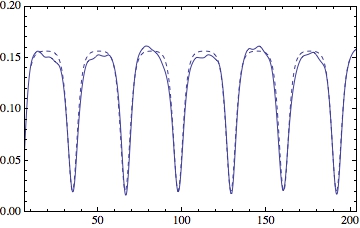,width = 0.95\columnwidth}\put(-240,127){ \textcolor{black}{$\rho(z)$}}\put(-95,2){ \textcolor{black}{$z$}}
\caption{ Details of analytical (dashed line) and numerically computed  (solid line)  density profiles $\left|\psi_{\rm 1 D}  (z,t) \right|^2$ of  dark soliton trains.  At time $t>0$ the interactions of the condensate on the left  are set to $g_L=3$ on the left-hand side of the domain in the numerical solution. The dimensionless units are used as specified in the main text.
\label{vergl}}
\end{figure}

We have compared the numerical generated soliton trains with the analytical periodic soliton trains due to the condensate wave functions $\eqref{appr}$. A typical example is presented in Fig. \ref{vergl} where parameters for the analytical solution where chosen to be as  follows. $v=0.1018$, $p=0.9978$, $n_0 =0.1566$ and $g=1$ while the numerical solution is considered on a space interval where self-interactions have been $g = 1$ for $t\le 0$ and $g_L=3$ for $t >0$.

A single bright soliton solution to \eqref{model1d} with constant attractive interactions is given by replacing $\tanh(z) \to 1/\cosh(z)$ in $\eqref{tanh}$ and setting $ g \to | g|$. The transition to the bright soliton train is obtained by $\tanh \to \operatorname{cn}$, i.e., 
\begin{multline}\label{appr2}
\psi_{\rm bt}(z,t)  = \sqrt{n_0} \Bigg[i  \frac{v}{c} + \sqrt{1- \left(\frac{v}{c}\right)^2  }\cdot \\  \cdot \operatorname{ cn } \Bigg(   \sqrt{ 1- \left(\frac{v}{c}\right)^2 } \cdot \sqrt{\frac{n_0 |g|}{2p^2}} \cdot \left(z -    v t \right) \bigg | p^2 \Bigg)  \Bigg].
\end{multline}
Again a straightforward calculation shows that it is approximately solving $\eqref{model1d}$, if $\mu$ is chosen to be $\mu= |g| n_0 \left( \frac{(1-2p^2)}{(2 p^2)}(1- \frac{v^2}{2n_0 g}) +  \frac{v^2}{2n_0 g} \right)$. Note that the chemical potential is negative for $1 >p^2 \gg 0$ and for $v \to 0$ and $\sqrt{p} \to 1$ converges to $\mu = - |g| n_0 /2$. Furthermore, this solution becomes exact either if  $\sqrt{p} \rightarrow 1$, where  $\operatorname{ cn }(z|1) = \frac{1}{\cosh(z)}$,  or if $v \rightarrow 0$, thereby generalizing the single bright soliton expression. We note that similar considerations  on the density profile like those made above for dark soliton train solutions apply to bright soliton train solutions as well. However, as the bright soliton train in Fig.~\ref{grain} is not freely expanding the limit $v \to 0$ gives the best approximation to our numerics.

  The densities corresponding to the dark soliton train solution $\eqref{appr}$ and the bright soliton train solution $\eqref{appr2}$ are related via
\beq\label{self}
\frac{ |\psi_{\rm dt}(z,t)|^2 +  |\psi_{\rm bt}(z,t)|^2}{\left(1 + \frac{v^2}{c^2} \right)} = n_0.
\eeq
\section{Emergence of soliton trains in quasi-one dimensional two-component BECs}
\label{sec-dark2}

In the previous sections we have found that the emergence and properties of soliton trains depend on the magnitude of change in self-interaction strength. Next we discuss how the state of a one-dimensional condensate of component $A$ $\psi^A_{\text 1D}$ is affected by the presence of a second component $B$ represented by $\psi^B_{\text 1D}$. Supposing $\omega \gg \omega_z$, rescaling time and neglecting trapping in $z$-direction the  wave functions are governed by the system
\begin{eqnarray}\label{uno1d}
 i \frac{\partial}{\partial t}  \psi^A_{\text 1D}= \left(-\frac{\partial^2}{\partial z^2}  + g_A |\psi^A_{\text 1D}|^2  + g_{AB} |\psi^B_{\text 1D}|^2 \right)\psi^A_{\text 1D}, \\
\label{due1d}
 i \frac{\partial}{\partial t}  \psi^B_{\text 1D}= \left(-\frac{\partial^2}{\partial z^2}   + g_B |\psi^B_{\text 1D}|^2 + g_{AB} |\psi^A_{\text 1D}|^2 \right)\psi^B_{\text 1D}.
\end{eqnarray}
Here the self-interactions $g_A$, $g_B$ and cross-interactions $g_{AB}$ are either constants or step functions. The dynamical stability of the mixture depends on the criterion $g_A g_B> g_{AB}^2$, therefore, by changing the interaction strength on one part of the cloud it is possible to have a miscible regime on one half and the phase separation regime on the other half of the domain.

\begin{figure}
\epsfig{file=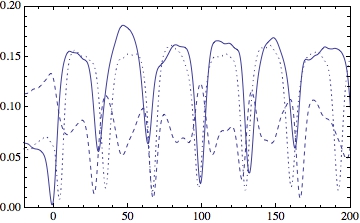,width=\columnwidth}\put(-253,129){ \textcolor{black}{$\rho(z)$}}\put(-92,2){ \textcolor{black}{$z$}}
\caption{   Density plot $\rho(z)=\left|\psi_{\rm 1 D}  (z,t) \right|^2$ of component $A$ of a coupled BEC (solid line), component $B$ (dashed line) at time $t=390$ and a single component condensate (dotted line). Initial state has $g_A=g_B=g_{AB}=1$. At time $t=0$ the interactions of the condensate A are set to $3$ on the left-hand side of the domain. The dimensionless units are used as specified in the main text. \label{hey}}
\end{figure}

First, we assume  that  initially ($t=0$) condensates $A$ and $B$ are spatially homogeneous with uniform and repulsive self- and cross-interactions. Then ($t=0^+$), self-interactions are changed in component $A$, i.e., formally $g_A \rightarrow g_A (z)$ leading to the generation of dark solitons and the appearance of complex dynamics for $t>0$. An example of the dynamics is shown in Fig.~\ref{hey} (with $g^L_A/g^R_A=3$ and $g_{AB}= g^R_A=g_B=1$). In that case, the soliton train in component $A$ is raised by the presence of the second condensate. Dark soliton trains generated in the presence of a second repulsive condensate are not as stable as single component condensates -  solitons decay faster.  However, we have observed dark soliton trains to appear at slightly lower interaction ratios than in single component condensates, which depends in particular on cross-interaction strength. 

%For example for $g^L/g^R=1.98$ and cross-interactions $g_{AB}=0.57$ dark soliton train appear.
%Soliton trains move faster than their single component counterpart, which might stem from the fact that they are raised by component $B$ implying an increase in the imaginary part of the wave function of component $A$ (see \eqref{appr}). 

 \begin{figure}
\epsfig{file= 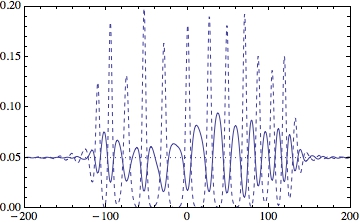,width=\columnwidth}\put(-253,129){ \textcolor{black}{$\rho(z)$}}\put(-98,2){ \textcolor{black}{$z$}}
\caption{  Snapshots of density plots $\rho(z)=\left|\psi_{\rm 1 d}  (z,t) \right|^2$ of a two component BEC at $t=0$ (dotted line) and $t=300$ - component $A$ (solid line) and $B$ (dashed line). Initial state has $g_A=1, g_B=-1, g_{AB}=1$. At time $t=0$ the interactions of the condensate A are set to $2.1$ on the left-hand side of the domain. The dimensionless units are used as specified in the main text. \label{mic}}
\end{figure}

  \begin{figure}
\epsfig{file=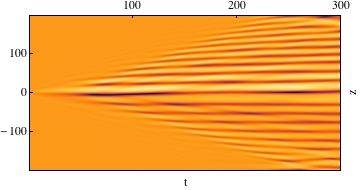,width=\columnwidth }
\caption{ Pseudo-color density plot $\rho(z)=\left|\psi_{\rm 1 D}  (z,t) \right|^2$ of component $A$ of a uniformly distributed Bose gas with constant interaction $g_A =g_{AB}=1=g^R$ and $g_B=-1$ at $t=0$ evolving in time as a change of $g^L/g^R =2.1$ has been implemented in $A$ for $t>0$. The change in interactions is sharp at $z=0$, i.e., $k \to \infty$ in $\eqref{step}$. Here luminosity is proportional to density. The dimensionless units are used as specified in the main text. \label{new}}
\end{figure}

 \begin{figure}
\epsfig{file=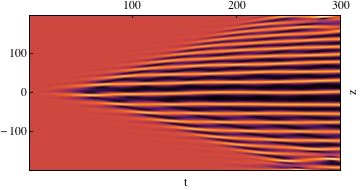,width=\columnwidth }
\caption{ Pseudo-color density plot $\rho(z)=\left|\psi_{\rm 1 D}  (z,t) \right|^2$ of component $B$ of a uniformly distributed Bose gas with constant interaction $g_B=-1$ and $g_A =g_{AB}=1=g^R$ at $t=0$ evolving in time as a change of $g^L/g^R =2.1$ has been implemented in $A$ for $t>0$. The change in interactions is sharp at $z=0$, i.e., $k \to \infty$ in $\eqref{step}$. Here luminosity is proportional to density. The dimensionless units are used as specified in the main text. \label{new2}}
\end{figure}

 Next we consider a  two component condensate where one component is attractive and the other component is repulsive. Initially both components are mixed and uniformly distributed. In Fig.~\ref{mic} snapshots of a two component condensate with initial parameters $g_A =1$, $g_B=-1$ are shown (dotted line). After changing self-interactions of component A to  $g^L_A/g^R_A=2.1$ a dark soliton train is generated. Fig.~\ref{new} and Fig.~\ref{new2} illustrate the spectrum of the time evolution for each component.  The dark soliton train in component $A$ represents the part of the effective potential for the other component $B$ and, therefore, induces excitations in condensate $B$ producing  a bright soliton train (dashed line).  As it can be seen in Fig.~\ref{mic} the density depletions of one component are at the maxima of the other and vice versa. In particular the frequency of the dark soliton train in $A$ is correlated with that in component $B$. As we showed above a change of self-interactions in an attractive single component condensate leads to a soliton train expanding in both directions (see Fig. \ref{brain2}). Hence, as the dark soliton train in $A$ for $z>0$ induces a bright soliton train in the other component $B$, which expands in both directions, this density depletion itself induces a dark soliton train expanding towards $z<0$ in component $A$ (Fig.~\ref{mic}.) Starting from the same initial distribution and changing interactions in the attractive condensate has a comparable effect, i.e., soliton trains in both components on the whole line are created. In any case only a very small change in self-interactions is sufficient to start this process, which is comparable to the behavior of the attractive single component condensate and due to its instability.

 General analytical solutions to two component condensates, where one component is in a state corresponding to a dark soliton train while the other component represents a bright soliton train can be constructed using the previous expressions $\eqref{appr}$ and $\eqref{appr2}$. Thus, a dark soliton train of the form
 \begin{multline}\label{apA}
\psi_{\rm A}(z,t)  = \sqrt{n_0} \Bigg[i  \frac{v}{c} + \sqrt{1- \left(\frac{v}{c}\right)^2  }\cdot \\  \cdot \operatorname{ sn } \Bigg(   \sqrt{ 1- \left(\frac{v}{c}\right)^2 } \cdot \sqrt{\frac{n_0 (g_A-g_{AB})}{2 p^2 }} \cdot \left(z -    v t \right) \bigg | p^2 \Bigg)  \Bigg],
\end{multline}
can be coupled to a bright soliton train of the form
\begin{multline}\label{apB}
\psi_{\rm B}(z,t)  = \sqrt{n_0} \Bigg[i  \frac{v}{c} + \sqrt{1- \left(\frac{v}{c}\right)^2  }\cdot \\  \cdot \operatorname{ cn } \Bigg(   \sqrt{ 1- \left(\frac{v}{c}\right)^2 } \cdot \sqrt{\frac{n_0 (g_B-g_{BA})}{2 p^2 }} \cdot \left(z -    v t \right) \bigg | p^2 \Bigg)  \Bigg],
\end{multline}
where one has to introduce appropriate chemical potentials in $\eqref{uno1d}$ and $\eqref{due1d}$ and both trains have the same periodicity.  We refer to appendix \ref{ApB} for a short outline.

\section{Controlled generation of vortex rings and soliton trains in $3D$}
\label{sec-vortex}

Let us now remove the constraint of one-dimensional geometries, but we still consider cigar-shaped traps. 
Due to the phenomenon of snake instability in dimensions higher than one, dark solitons in repulsive condensates decay into more stable excitations such as vortices \cite{Brand, berloffRP} or vortex rings. Thus we would expect that once dark solitons are generated, they would decay into vortex rings in three-dimensions and the threshold in the self-interaction imbalance for the generation of these excitations would be close to the one obtained in the quasi-one dimensional system discussed earlier.

The scenario of vortex rings nucleation is very similar to that of vortex rings formation after a cavity collapse \cite{berloffBarenghi}. When the train of shock waves/dark solitons is formed, some part of the front breaks into vortex rings with an extra energy necessary  to drive such transition provided by the part of the train traveling behind \cite{pade}.  The energy transfer also counteracts the effect of the friction allowing the ring to travel a long distance before breaking apart.

The presence of the solitary wave  train enhances the instability leading to the formation of vortex rings in comparison with the instability of a single grey soliton. The faster the soliton moves the more stable it becomes. To overcome this stability there has to exist the supply of energy which is provided by the waves traveling behind.

%Let us first turn to the phenomenon of vortex rings in a $3D$ cylindrically symmetric setting. The proposed mechanism to generate these excitations is the same as for the generation of dark soliton trains, i.e. for a single component condensate, drastically changing self-interactions on one side of the condensate (see Fig. \ref{post}). It should be pointed out that there is a deeper connection between the phenomenon of dark soliton trains in $1D$ and  the emergence phenomena of vortices or vortex rings in higher dimensions. Indeed in dimensions $2D$ and $3D$ dark solitons decay due to the phenomenon of snake instability into more stable excitations such as vortices \cite{Brand} or vortex rings and could thereby be regarded as the lower dimensional sister of vortex rings in a $3D$ spherically symmetric trap. Hence, one should expect both nonlinear excitations to appear at a similar change in self-interactions.

%%%%%%%%%%%%%%%%%%%%%%%%%%%%%%%%%
 
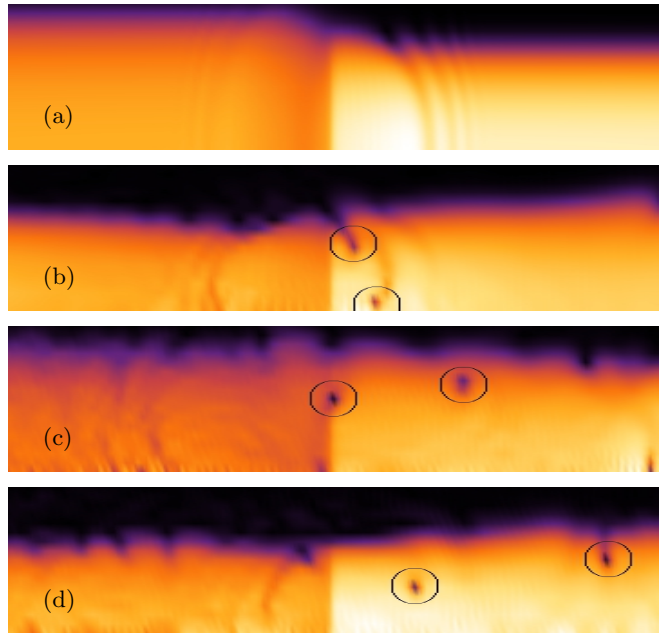
\begin{figure}\hbox{
\begin{picture} (100,60)
 \put(0,0){\includegraphics[width=\columnwidth, height=55pt]{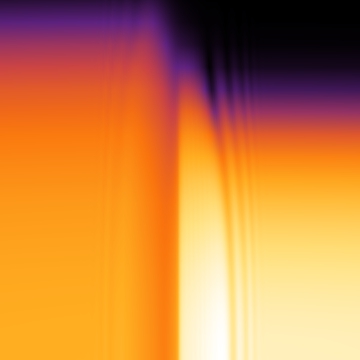} } \put(10,10){ \textcolor{black}{(a)}}
\end{picture} }
\hbox{ \begin{picture} (100,60)
 \put(0,0){\includegraphics[width=\columnwidth, height=55pt]{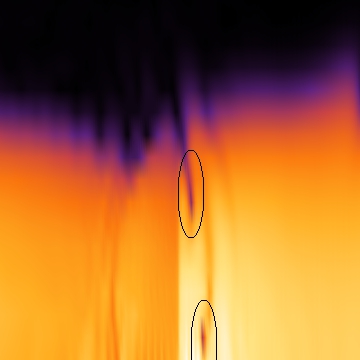} } \put(10,10){ \textcolor{black}{(b)}}
\end{picture}  }
\hbox{ \begin{picture} (100,60)
 \put(0,0){\includegraphics[width=\columnwidth, height=55pt]{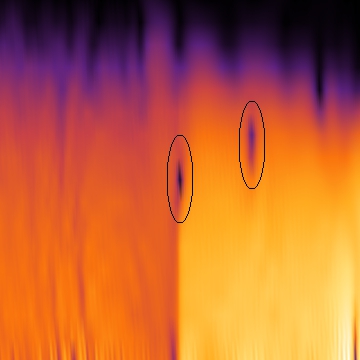} } \put(10,10){ \textcolor{black}{(c)}}
\end{picture}  }
\hbox{ \begin{picture} (100,60)
 \put(0,0){\includegraphics[width=\columnwidth, height=55pt]{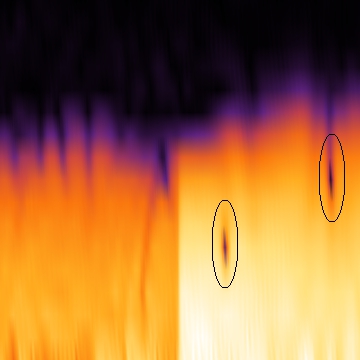} } \put(10,10){ \textcolor{black}{(d)}}
\end{picture}  }

\caption{(Color online) Same as in Fig.~\ref{ex2} but for $g^L/g^R =2.1$ and times (a) $t=0.75$, (b) $t=4.5$, (c) $t=7.5$, (d) $t=11.25$. The spatial region shown corresponds to $z\in[-20,20], r\in [0,8]$. Black corresponds to low atom densities and yellow to high ones. The dimensionless units are used as specified in the main text. \label{ex2}}
\end{figure}

\begin{figure}
\hbox{
\begin{picture} (100,60)
 \put(0,0){\includegraphics[width=\columnwidth, height=55pt]{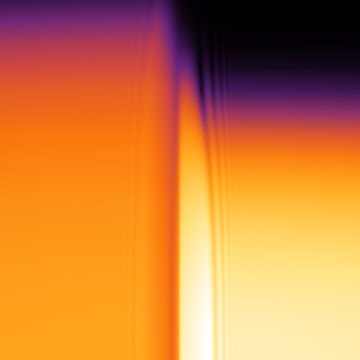} } \put(10,10){ \textcolor{black}{(a)}}
\end{picture} }
\hbox{ \begin{picture} (100,60)
 \put(0,0){\includegraphics[width=\columnwidth, height=55pt]{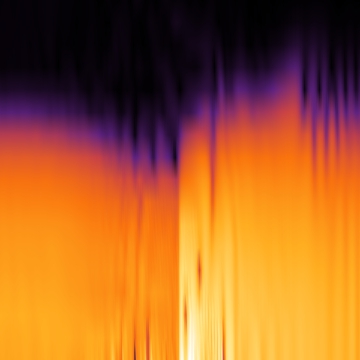} } \put(10,10){ \textcolor{black}{(b)}}
\end{picture}  }
\hbox{ \begin{picture} (100,60)
 \put(0,0){\includegraphics[width=\columnwidth, height=55pt]{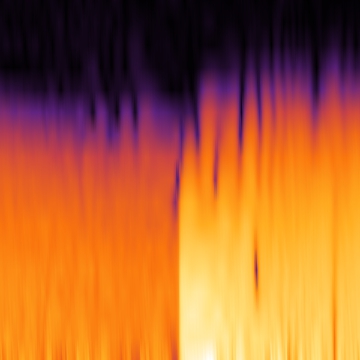} } \put(10,10){ \textcolor{black}{(c)}}
\end{picture}  }
\hbox{ \begin{picture} (100,60)
 \put(0,0){\includegraphics[width=\columnwidth, height=55pt]{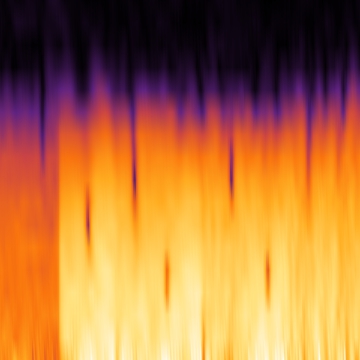} } \put(10,10){ \textcolor{black}{(d)}}
\end{picture}  }
\caption{(Color online) Pseudocolor plot of atom density $|\psi(r,z,t)|^2$ snapshots for different values of time: (a) $t=0.75$, (b) $t=4.5$,  (c) $t=8.625$, (d)  $t=13.875$ and 
(e) $t=18.375$. The values of the interactions $g^L/g^R =2.3$ and the spatial region shown corresponds to $z\in[-30,30], r\in [0,8]$ in subplots (a-d) and to
 $z\in [-10,50]$ in subplot (e). Black corresponds to low atom densities and yellow to high ones.  The dimensionless units are used as specified in the main text. \label{ex3}}
\end{figure}
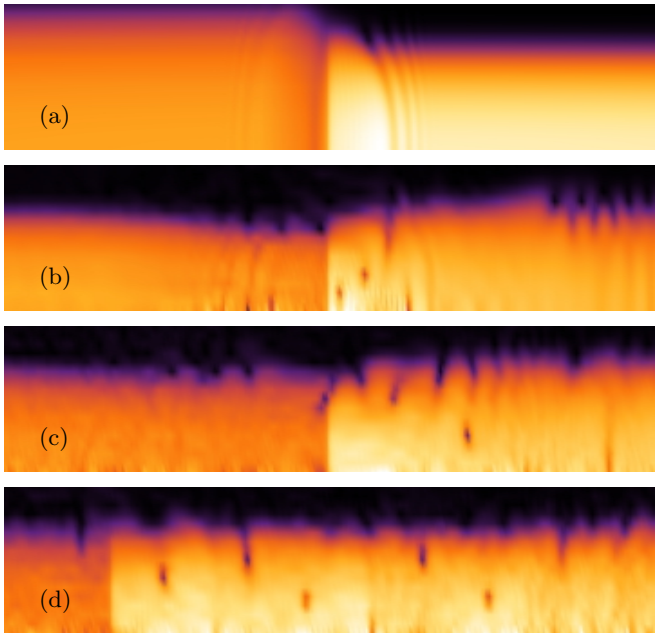

%%%%%%%%%%%%%%%%%%%%%%%%%%%%%%%

\subsection{Dynamics of single component BEC}

We have numerically simulated Eq.~(\ref{model}) for various parameter combinations \cite{mynote2} and will
%\begin{figure}\hbox{
%\begin{picture} (100,60)
% \put(0,0){\includegraphics[width=\columnwidth, height=55pt]{nn1.jpg} } \put(10,10){ \textcolor{black}{(a)}}
%\end{picture} }
%\hbox{ \begin{picture} (100,60)
% \put(0,0){\includegraphics[width=\columnwidth, height=55pt]{nn3.jpg} } \put(10,10){ \textcolor{black}{(b)}}
%\end{picture}  }
%\hbox{ \begin{picture} (100,60)
% \put(0,0){\includegraphics[width=\columnwidth, height=55pt]{nn2.jpg} } \put(10,10){ \textcolor{black}{(c)}}
%\end{picture}  }
%\hbox{ \begin{picture} (100,60)
% \put(0,0){\includegraphics[width=\columnwidth, height=55pt]{nn4.jpg} } \put(10,10){ \textcolor{black}{(d)}}
%\end{picture} }
%\caption{(Color online) Pseudocolor plot of atom density $|\psi(r,z,t)|^2$ snapshots for different values of time: (a) $t=1.5$, (b) $t=9$,  (c) $t=10.5$ and (d)  $t=15$. 
%The values of the interactions $g^L/g^R =2.05$ and the spatial region shown corresponds to $z\in[-30,30], r\in [0,8]$. Black corresponds to low atom densities and yellow to high ones. \label{ex1}}
%\end{figure}
describe the typical outcome for a specific example corresponding to a large repulsive BEC with $g = 10^5$, with $\lambda_x=\lambda_y=1, \lambda_z = 0.05$ (i.e. soft longitudinal trapping). Our initial configuration is a ground state BEC corresponding to $ g=g^L=g^R$. At time $t=0$ we suddenly raise interactions strength $g_L$  for $z<0$  and then observe the subsequent evolution of the condensate.

%The results of a series of simulations to be described in detail later are summarized in Fig. \ref{prima}. 
Once the non-equilibrium situation is generated there is a flow of atoms from $z<0$ to $z>0$  with a flow intensity depending on the ratio $g^L/g^R$. When a critical value $g^L/g^R \simeq 2$ is surpassed the vortex rings are generated as seen on  Fig.~\ref{ex2}. 
% For $g^L/g^R =2.05$  just above the critical value Fig.  \ref{ex1} illustrated the stages in vortex ring formation. For small times (Fig.  \ref{ex1} (a)) the sudden increase of interactions results in a strong flow of atoms coming from the region with $z<0$ towards the region with $z>0$ (right) resulting in the formation of a shock wave \cite{V2}. Shortly after that, a small vortex ring appears in the low density regions around $z=0$ but it counterflows towards $z<0$ and disappears in that low density unstable region, see Fig.  \ref{ex1}(b)-(c). After that, the dark soliton evolves into  a single vortex at $z>0$, see Fig.  \ref{ex1}(d).
%\begin{figure}
%\input{Vv3.tex}
%\caption{(Color online) Pseudocolor snapshots  of  density $|\psi(r,z,t)|^2$  for  $g^L/g^R =2.1$ and times (a) $t=1.5$, (b) $t=9$, (c) $t=15$, (d) $t=22.5$. The spatial region shown corresponds to $z\in[-20,20], r\in [0,8]$. Black corresponds to low atom densities and yellow to high ones.  Circles in dicate the cross sections of vortex rings.  \label{ex2}}
%\end{figure}
In Fig.~\ref{ex2} we present the stages of vortex rings formation for  $g^L/g^R =2.1$.  The change in interaction strength leads to a generation of  a  train  of dark solitons, see Fig.~\ref{ex2}(a),  that evolve into vortex rings which enter the condensate around $z=0$ coming from the low density region, see Fig. ~\ref{ex2}(b),  and move slowly through the condensate remaining stable for long times  [Fig.~\ref{ex2}(c,d)]. Another vortex ring with a large radius seems to be present in the lower density regions where it would be experimentally difficult to detect.

%\begin{figure}
%\hbox{
%\begin{picture} (100,60)
% \put(0,0){\includegraphics[width=\columnwidth, height=55pt]{kk1.jpg} } \put(10,10){ \textcolor{black}{(a)}}
%\end{picture} }
%\hbox{ \begin{picture} (100,60)
 %\put(0,0){\includegraphics[width=\columnwidth, height=55pt]{kk2.jpg} } \put(10,10){ \textcolor{black}{(b)}}
%\end{picture}  }
%\hbox{ \begin{picture} (100,60)
% \put(0,0){\includegraphics[width=\columnwidth, height=55pt]{kk3.jpg} } \put(10,10){ \textcolor{black}{(c)}}
%\end{picture}  }
%\hbox{ \begin{picture} (100,60)
% \put(0,0){\includegraphics[width=\columnwidth, height=55pt]{kk4.jpg} } \put(10,10){ \textcolor{black}{(d)}}
%\end{picture}  }
%\caption{(Color online) Pseudocolor plot of atom density $|\psi(r,z,t)|^2$ snapshots for different values of time: (a) $t=0.15$, (b) $t=9$,  (c) $t=17.25$, (d)  $t=27.75$ and 
%(e) $t=36.75$. The values of the interactions $g^L/g^R =2.3$ and the spatial region shown corresponds to $z\in[-30,30], r\in [0,8]$ in subplots (a-d) and to
% $z\in [-10,50]$ in subplot (e). Black corresponds to low atom densities and yellow to high ones. \label{ex3}}
%\end{figure}

Increasing the interactions even further to $g^L/g^R =2.3$ leads to a richer dynamics as summarized in Fig.~\ref{ex3}. The short-time dynamics is analogous to the previous cases [Fig.~\ref{ex3}(a)] but then a complex transient appears where several vortex rings enter the condensate; also rarefaction pulses are clearly identified [see Fig.~\ref{ex3}(b)]. After that, some of those vortices counter-flow and disappear and a much more regular picture arises with several vortex rings moving to the right in a very clear way. Fig. \ref{ex3}(c) shows few vortex rings slowly moving through the condensate for $t=8.625$ and one being generated around $z=0$. Fig.~\ref{ex3}(d) shows a later stage of the evolution where three long-lived vortex rings travel smoothly through the condensate although their relative positions changes due to differences in their speeds (notice the small differences in their radii) and their interaction with  sound waves originated after the reflection of the shock wave in the condensate boundary [see Fig. \ref{ex3}(e)].
% The increasing number of vortex rings generated as higher change in interactions is implemented is in accordance to the relation between frequencies of dark soliton trains and change in interactions in $1D$ systems. In opposition to this relation however stands that a higher change is accompanied by stronger fluctuations within the condensate that lead to a faster decay of vortex rings.

\begin{figure}
\hbox{
\begin{picture} (100,60)
 \put(0,0){\includegraphics[width=\columnwidth, height=55pt]{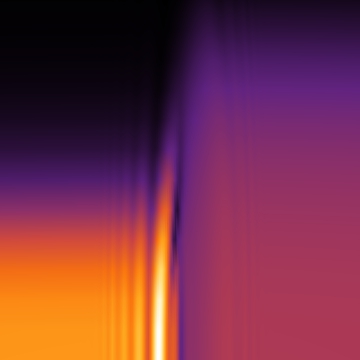} } \put(10,10){ \textcolor{black}{(a)}}
\end{picture} }
\hbox{ \begin{picture} (100,60)
 \put(0,0){\includegraphics[width=\columnwidth, height=55pt]{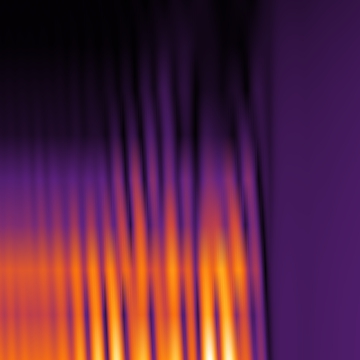} } \put(10,10){ \textcolor{black}{(b)}}
\end{picture}  }
\caption{(Color online) Pseudocolor plot of atom density $|\psi(r,z,t)|^2$ snapshots for different values of time: (a) $t=0.75$ and (b) $t=3.75$. The values of the interactions are $g^L/g^R = -10^{-3}$. Black corresponds to low atom densities and yellow to high ones. The spatial region shown corresponds to $z\in[-20,20], r\in [0,6]$ in subplot (a) and to
 $z\in [-30,10], r\in [0,4]$ in subplot (b). The dimensionless units are used as specified in the main text. \label{attr1}}
\end{figure}
Fig. \ref{attr1} shows that changing interactions to attractive, i.e., a change of $g^L/g^R =-10^{-3}$ for $z<0$ causes the formation of a shock wave there and the formation of solitonic waves,  similar to the formation of bright soliton trains in quasi-one dimensional BECs.  Due to the small attractive force between particles  and the tight potential in the transverse direction the condensate gently collapses into a quasi-one dimensional setup, thereby carrying stable solitons in its attractive part. Increasing the attractive force to more negative values of scattering length leads to a blowup of the condensate wave function, while lowering implies more stable solitary waves even for BEC without interaction ($g^L=0$). Lowering  scattering lengths on the l.h.s. to positive values leads to vortex ring generation once a threshold is surpassed and to solitary waves as $g^L$ tends to become smaller.

\subsection{The two component case}

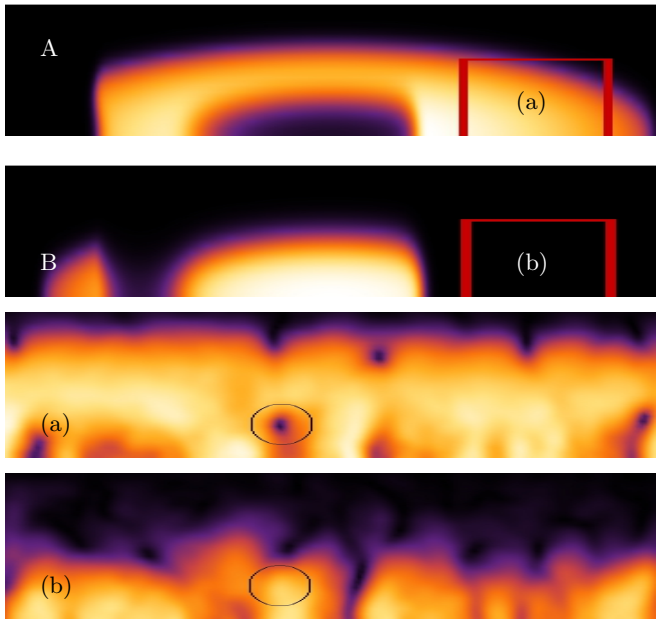
\begin{figure}
\hbox{
\begin{picture} (100,60)
 \put(0,0){\includegraphics[width=\columnwidth, height=55pt]{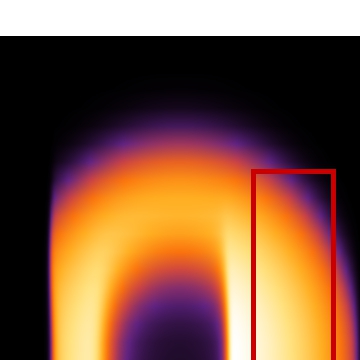} } \put(190,10){ \textcolor{black}{(a)}} \put(10,30){ \textcolor{white}{A}}
\end{picture} }
\hbox{ \begin{picture} (100,60)
 \put(0,0){\includegraphics[width=\columnwidth, height=55pt]{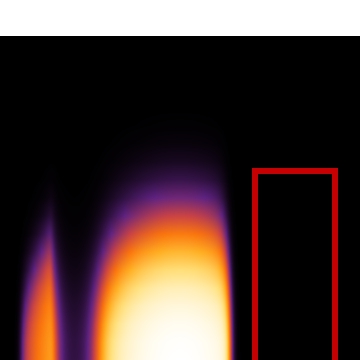} } \put(190,10){ \textcolor{white}{(b)}} \put(10,10){ \textcolor{white}{B}}
\end{picture}  }
\hbox{ \begin{picture} (100,60)
 \put(0,0){\includegraphics[width=\columnwidth, height=55pt]{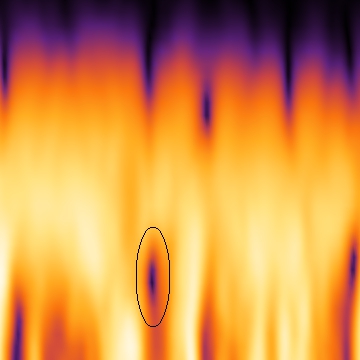} } \put(10,10){ \textcolor{black}{(a)}}
\end{picture}  }
\hbox{ \begin{picture} (100,60)
 \put(0,0){\includegraphics[width=\columnwidth, height=55pt]{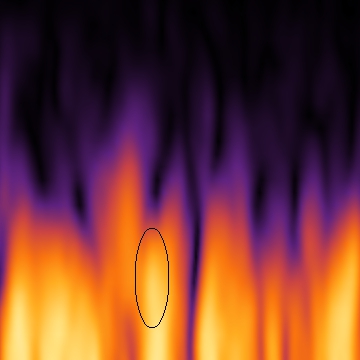} } \put(10,10){ \textcolor{black}{(b)}}
\end{picture}  }
\caption{ (Color online) Pseudocolor plot of atom density $|\psi(r,z,t)|^2$ snapshots of component $A$ and component $B$ in the initial state at $t=0$. The spatial region shown corresponds to
 $z\in [-120,120]$ and $r\in [0,10]$. Subplots (a),(b) at $t=35.25$ show details for a region  $z\in [30,60]$, $r\in [0,7]$. The circle marks the position of a ``skyrmion" -- vortex in the first component with the density maximum of the second component. The dimensionless units are used as specified in the main text.\label{skyr}}
\end{figure}

 We now turn to two component systems of clearly distinct states $\psi_A$ and $\psi_B$ with interactions to be chosen within the phase separation  regime, $g_{AB}^2 > g_A g_B$, i.e. cross-interactions between both components are dominating. The harmonic trapping potential for the repulsive two component BEC has been specified by $\lambda_x=\lambda_y=1, \lambda_z = 0.05$ (i.e. soft longitudinal trapping) for both components. To generate a quantum piston induced evolution of the condensate wave functions containing skyrmions  we tested various different initial conditions.  The initial ground state at $t=0$ on which we apply the quantum piston scheme consists of one component surrounded by the second component. The initial states are naturally generated by putting both components (each localized regarding a harmonic trap specified by  $\lambda_x=\lambda_y=1, \lambda_z = 0.05$ but one translated along the $z$-axis to the left and the other to the right) into the single trap without any overlapping of the atom clouds and by evolving  the corresponding state in imaginary time until the new common ground state is reached at $t=0$.

In Fig. \ref{skyr} we show an example of density profiles of a two component BEC in such a ground state at $t=0$, which is specified by its self-interactions $g_A = 6005$, $g_B=2650$ and cross-interactions $g_{AB}=6000$. After a change in self-interactions by a factor $g^L_A/g^R_A=2.26$ on the l.h.s. in component $A$ has been implemented vortices are generated. In particular we observe the emergence of a vortex ring in component $A$ that is filled with mass of component $B$, which can be identified as  a skyrmion - the corresponding area within the density distributions in Fig. \ref{skyr} is encircled. 

\section{Conclusions}
\label{sec-conclusions}
  In this paper we have studied several examples of how the spatial and temporal control of the self-interactions in an atomic BEC leads to the formation of  nonlinear excitation such as dark/bright solitons/ solitary trains and solitary waves using a nonlinear ``quantum piston" concept. In an axisymmetric elongated condensate vortex rings form as the interaction strength on one half of the condensate changes by a factor exceeding 2. This mechanism can be used to controllably generate and study such excitations. 
Our proposal, in addition to being conceptually simple and accessible to present experimental techniques improves essentially 
currently used methods to produce nonlinear excitations.  Two component Bose-Einstein condensates could be used to amplify the generation of vortex rings or solitons and give rise to another set of excitations  such as skyrmions. 
The number of vortex rings/skyrmions generated can be controlled by  changing the transverse confinement $\omega/\omega_z$. For weak transverse confinement a moving solitary wave is subject to snake instability leading to the formation of vortex rings. For a sufficiently tight transverse confinement the solitary wave becomes stable to the snake instability and vortices will not form. 

The faster a solitary wave moving the more stable it becomes to the snake instability. In the periodic train the stability is reduced because of the energy transfer between the parts of the train \cite{pade}. The vortex generation in the proposed method, therefore, is the result of an intricate interplay between shock wave train generation and the  snake instability enhanced by the energy transfer between the parts of the train.
%We believe that future research on our method of controlled excitation generation  could lead to a simple realization of controlled transport of subatomic particles within quantum fluids by using those excitations as vehicles for subatomic particles.

\acknowledgments

 F.P. was partly supported by EPSRC and Peter Markowich's KAUST grant and his work was carried out at the Cambridge Center for Analysis (CCA). N. G. B. acknowledges support from FP7 CLERMONT4  PITNGA-2009235114. V. M. P.-G. is partially supported by grant  MTM2012-31073  (Ministerio de Econom\'{\i}a y Competitividad, Spain). 

\appendix
  \section{Stationary solutions to GPE with step-like coupling parameter}
  \label{solutions2}
  
We will describe here a procedure to construct time independent solutions of the GPE \eqref{model1d}  with step function coupling parameters. To introduce the method we will consider the second derivative of a `two branches of the real line' ansatz defined as
\beq\label{twob}
u(z) =  \tanh(g_{+} (z)) + \tanh(g_{-} (z)).
\eeq
Here the basic idea is that one branch (denoted by the subscript $-$) takes into account properties of the solution for $z<0$ and the other branch (denoted by the subscript $+$) properties relevant in $z>0$. The arguments of the hyperbolic tangent, $g_+$ and $g_-$, are explicitly given by
\beq\label{gee}
g_{+}(z) =  \lim_{k \to \infty}\frac{\log \left(e^{2 k z} +1 \right)}{2 k} =  \begin{cases} |z|  & z > 0 \\ 0 &  \text{otherwise} \end{cases}
\eeq
and
\beq
g_{-}(z) = \lim_{k \to \infty}\frac{- \log \left(e^{- 2 k z} +1 \right)}{2 k} =  \begin{cases} 0  & z \geq 0 \\ -  |z| &  \text{otherwise} \end{cases}
\eeq
Consequently the derivatives of  these functions are
\beq
g'_{+}(z) =  \lim_{k \to \infty} \frac{1}{1+ e^{- 2 k z}} =  \begin{cases} 1  & z > 0 \\ 1/2  & z = 0 \\ 0 & \text{otherwise} \end{cases}
\eeq
and
\beq
g'_{-}(z) = \lim_{k \to \infty} \frac{1}{1+ e^{2 k z}} =  \begin{cases} 1  & z < 0 \\ 1/2 & z = 0 \\ 0  &  \text{otherwise} \end{cases}
\eeq
and the second derivatives satisfy
\beq
g''_{+}(z) =  \begin{cases} 0  & z \neq 0 \\ \infty  & z = 0 \end{cases}
\eeq
and
\beq
g''_{-}(z) =  \begin{cases} 0  & z \neq 0 \\ - \infty  & z = 0 \end{cases}.
\eeq
These functions obey the property
\beq\label{bye}
- g''_{+}(0) = g''_{-}(0)  =  \lim_{k \to \infty} \left(- \frac{k}{2} + k \right) =  \infty .
\eeq
Now consider the second derivative of our ansatz, i.e., 
\begin{multline}
\partial_{z}^2 u(z) = \text{sech}^2(g_{+} ) g_{+}''   + \text{sech}^2(g_{-}) g_{-}''  - \\Ê-  2 \text{sech}^2(g_{+} ) \tanh(g_{+}) g_{+}'^2   - 2\text{sech}^2(g_{-} ) \tanh(g_{-}) g_{-}'^2,
\end{multline}
at $z = 0$, i.e., where by $\eqref{bye}$ and the fact that $g_+(0) =g_-(0) = 0$ one gets
 \beq
 \text{sech}^2(g_{+} ) g_{+}''   + \text{sech}^2(g_{-}) g_{-}''  =  (g_{+}''   +g_{-}''  ) = 0.
 \eeq
Hence the second derivative of $\eqref{twob}$ satisfies
 \begin{multline}
 \partial_{z}^2 u(z=0) =   2  \left(\tanh(g_{+})^2 -1  \right)\tanh(g_{+}) g_{+}'^2  + \\ + 2  \left(\tanh(g_{-})^2 -1  \right)\tanh(g_{-}) g_{-}'^2 = 0,
\end{multline}
and 
 \begin{multline}
  \partial_{z}^2 u(z\neq 0) =   2  \left(\tanh(g_{\pm})^2 -1  \right)\tanh(g_{\pm}) g_{\pm}'^2,
\end{multline}
where the subindexes  $+$ and $-$ correspond to $z > 0$ and $z <0$ respectively

Let us go a step further and rescale our ansatz in order to get a solution for
\beq\label{kaka}
\partial_z^2 u  = \theta(z) |u|^2 u - \mu u
\eeq
 with
 \beq
  \theta(z) = \begin{cases} g_1 \hspace{9mm} \text{ if }   z  < 0\\  c \hspace{11mm}  \text{ if }   z  =  0\\  g_2  \hspace{9mm} \text{ if }   z > 0. \end{cases}
 \eeq
The rescaled two branches ansatz is given by
\begin{multline}\label{na1}
u\equiv u_+  + u_-  \equiv \sqrt{\frac{\mu}{g_2}} \tanh \left[  \pm \sqrt{ \frac{\mu}{2}} \tilde g_+(z)   \right] + \\  \sqrt{\frac{\mu}{g_1}} \tanh \left[  \pm \sqrt{ \frac{\mu}{2}} \tilde g_-(z)   \right] 
\end{multline}
with phase functions defined by
\beq
\tilde g_{+}(z) =  \lim_{k \to \infty}\frac{\log \left(e^{2 \sqrt{g_1} k z} +1 \right)}{2 \sqrt{g_1} k} =  \begin{cases} |z|  & z > 0 \\ 0 & \text{otherwise} \end{cases}
\eeq
and
\beq
\tilde g_{-}(z) = - \lim_{k \to \infty}\frac{\log \left(e^{- 2 \sqrt{g_2} k z} +1 \right)}{2 \sqrt{g_2} k} =  \begin{cases} 0  & z \geq 0 \\ -  |z|  & \text{otherwise} \end{cases}
\eeq
\begin{figure}
\epsfig{file=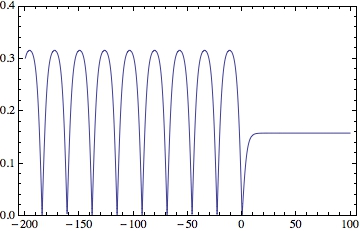,width = 0.95\columnwidth}\put(-245,129){ \textcolor{black}{$\rho(z)$}}\put(-97,2){ \textcolor{black}{$z$}}
\caption{   \label{snake} Density plot $\left|\psi_{\rm 1 d}  (z,t) \right|^2$ of an example of an exact two branch  real line solution to the GPE with constant self-interactions. \label{ana}}
\end{figure}
To verify that  \eqref{na1} has the desired property we consider its second derivative 
 \begin{multline}\label{klo}
\partial_{z}^2 \left( u_+ + u_-  \right)= \frac{\mu}{\sqrt{ 2g_2}}  \text{sech}^2(g_{+} ) g_{+}''   + \frac{\mu}{\sqrt{ 2g_1}} \text{sech}^2(g_{-}) g_{-}''  -  \\ -  \mu^{3/2} \bigg(\frac{1}{\sqrt{g_2}} \text{sech}^2(g_{+} ) \tanh(g_{+}) g_{+}'^2 - \\  - \frac{1}{\sqrt{g_1}} \text{sech}^2(g_{-} ) \tanh(g_{-}) g_{-}'^2 \bigg)  \\ = g_2 u_+^3 - \mu u_+ + g_1 u_-^3 - \mu u_-.
\end{multline}
Note that $u_+$ is nonzero, iff $z >0$, as well as $u_-$ is nonzero, iff $z <0$. 

Furthermore one can interchange one branch or both by a Jacobi elliptic type function of a similar form as the hyperbolic tangent branches, which is a solution to the same differential equation $\eqref{kaka}$.  In Fig. \ref{ana} one finds an example of such a solution. On the left hand side one finds the Jacobi elliptic part while on the r.h.s. the density corresponds to a hyperbolic tangent wave function.

\section{Determining self-interaction strength $g$ of the condensate via the form of the soliton train}
\label{ApX}

The `free' parameters of the condensate wave function $\eqref{appr}$ are $\{n_0 ,v,p,g \}$ and
the form of the analytical dark soliton train solution depends on the parameter $p$ in $\eqref{appr}$ - if $p$ is close to $1$ the contribution of the hyperbolic tangens is dominating, while for smaller $p$ the solution resembles properties of a squared sinus. Hence, one selects the elliptic modulus $p$ by comparing the form of the numerically generated profile with the form of the analytical expression. The amplitude of the solution and the depth of each soliton are fixed by the requirement  that at a maximum of the density graph we have
\beq\label{e1}
 \max \big| \psi_{\rm dt} \big|^2 = n_0 = c_1,
\eeq
and at a minimum
\beq\label{e2}
 \min \big| \psi_{\rm dt} \big|^2 = \left( \frac{v}{\sqrt{2 g}} \right)^2 = c_2,
\eeq
where $c_1$ and $c_2$ are fixed numbers. The periodicity is fixed as well, i.e.,
\beq\label{e3}
 \sqrt{ 1- \left(\frac{v}{c}\right)^2 } \cdot \sqrt{\frac{n_0 g}{2 p^2}} = c_3,
\eeq
where $c_3$ again is a constant. Inserting $\eqref{e1}$ and $\eqref{e2}$ in $\eqref{e3}$ determines $g$.  Hence we can determine the effective interactions between atoms and  the velocity of the dark soliton train  from the density profile of the condensate at a particular instant in time.

\section{Dark-bright soliton train solutions for two component BEC}
\label{ApB}

 We now show that there exist analytical expressions for a dark soliton train in component $A$ coupled to a bright soliton train in component $B$. We recognize that
\begin{multline}\label{apAA}
\psi_{\rm A}(z,t)  = \sqrt{n_0} \Bigg[i  \frac{v}{c} + \sqrt{1- \left(\frac{v}{c}\right)^2  }\cdot \\  \cdot \operatorname{ sn } \Bigg(   \sqrt{ 1- \left(\frac{v}{c}\right)^2 } \cdot \sqrt{\frac{n_0 (g_A-g_X)}{2 p^2 }} \cdot \left(z -    v t \right) \bigg | p^2 \Bigg)  \Bigg],
\end{multline}
satisfies the equation
\begin{equation}\label{herr}
 i  \partial_t  \psi_{A}= \left( - \partial_z^2 +  \left(g_A |\psi_{A}|^2 - g_{X} |\psi_{A}|^2 - \mu \right) \right)\psi_{A}.
\end{equation}
We rewrite some terms
\begin{multline}
 - g_{X} |\psi_{A}|^2 - \mu = - g_{X} n_0 \left(\frac{v^2}{c^2} + \left(1- \frac{v^2}{c^2} \right) \operatorname{sn}^2 \right) - \mu  = \\ =
 g_{X} n_0 \left( \frac{v^2}{c^2} +\left(1- \frac{v^2}{c^2} \right) \operatorname{cn}^2 \right) - \tilde \mu
\end{multline}
with $\tilde \mu = 2 g_X n_0 \frac{v^2}{c^2} + g_X n_0 \left(1 - \frac{v^2}{c^2} \right) + \mu$. Hence, by setting $g_{X} \to g_{AB}$ and defining
\begin{multline}\label{apBB}
\psi_{\rm B}(z,t)  = \sqrt{n_0} \Bigg[i  \frac{v}{c} + \sqrt{1- \left(\frac{v}{c}\right)^2  }\cdot \\  \cdot \operatorname{ cn } \Bigg(   \sqrt{ 1- \left(\frac{v}{c}\right)^2 } \cdot \sqrt{\frac{n_0 (g_B - g_Y)}{2p^2}} \cdot \left(z-    v t \right) \bigg | p^2 \Bigg)  \Bigg],
\end{multline}
with $g_B - g_Y = c = g_A - g_{AB} > 0$
$\eqref{apAA}$ satisfies
\begin{equation}\label{unsinn}
i  \partial_t  \psi_{A}= \left( - \partial_z^2 +  \left(g_A |\psi_{A}|^2 + g_{AB} |\psi_{B}|^2 - \tilde \mu \right) \right)\psi_{A}.
\end{equation}
On the other hand $\eqref{apBB}$ satisfies
\begin{multline}\label{unsinn2}
 i  \partial_t  \psi_{B}= \left( - \partial_z^2 +  \left( (g_{B}-g_{Y}) |\psi_{B}|^2 - \mu' \right) \right)\psi_{B} = \\ = \left( - \partial_z^2 +  \left( g_{B} |\psi_{B}|^2  + g_{AB} |\psi_{A}|^2 - \mu'' \right) \right)\psi_{B},
\end{multline}
by setting $g_{Y} \to g_{AB}$ and for an appropriately chosen $\mu''$.


\begin{thebibliography}{10}
\bibitem{mynote}
 In simulations, we used $\Delta x = 1/6$, $\Delta t=0.0015$,  $z\in (-500,500)$, $\mu=0.1$ and reflective boundary conditions.
\bibitem{mynote2}
We used a computational window of $240$ space units in $z$ and $10$ units along the transverse directions $x,y$. Typical space steps are about $\Delta x=0.25$ and the time steps $\Delta t \simeq 1.5\times 10^{-4}$.
\end{thebibliography}
  \end{document}